\documentclass{elsart}
\usepackage{graphicx}
\usepackage{epsfig}
\usepackage{amssymb}
\usepackage{color}

\begin{document}
 \begin{frontmatter}

 \title{Monte Carlo study of multicomponent monolayer adsorption on square lattices}

\author[label1,label2]{G. D. Garc\'{\i}a},
\author[label1,label2]{F. O. S\'anchez-Varretti},
\author[label1]{F. Bulnes},
\author[label1]{A. J. Ramirez-Pastor\corauthref{cor1}}

\address[label1]{Dpto. de F\'{\i}sica, Instituto de F\'{\i}sica Aplicada, Universidad Nacional de San Luis - CONICET, Chacabuco 917, 5700 San Luis, Argentina.}
\address[label2]{Universidad Tecnol\'ogica Nacional, Regional San Rafael, Gral. Urquiza 314, 5600, San Rafael, Mendoza, Argentina. }
\thanks[cor1]{Corresponding author. Fax +54-2652-430224, E-mail: antorami@unsl.edu.ar}

\begin{abstract}
The monolayer adsorption process of interacting binary mixtures of
species $A$ and $B$ on square lattices is studied through grand
canonical Monte Carlo simulation in the framework of the
lattice-gas model. Four different energies have been considered in
the adsorption process: 1) $\epsilon _0$, interaction energy
between a particle (type $A$ or $B$) and a lattice site; 2)
$w_{AA}$, interaction energy between two nearest-neighbor $A$
particles; 3) $w_{BB}$, interaction energy between two
nearest-neighbor $B$ particles; 4) $w_{AB}=w_{BA}$, interaction
energy between two nearest-neighbors being one of type $A$ and the
other of type $B$. The adsorption process has been monitored
through total and partial isotherms and differential heats of
adsorption corresponding to both species of the mixture. Our main
interest is in the repulsive lateral interactions, where a variety
of structural orderings arise in the adlayer, depending on the
interaction parameters ($w_{AA}$, $w_{BB}$ and $w_{AB}$). At the
end of this work, we determine the phase diagram characterizing
the phase transitions occurring in the system. A nontrivial
interdependence between the partial surface coverage of both
species is observed.
\end{abstract}

\end{frontmatter}


\newpage

\section{Introduction}

The adsorption process of mixture gases on solid surfaces is a
topic of great interest not only from an intrinsic but also from a
technological point of view, due to its importance for new
developments in fields like gas separation and purification
\cite{Ruthven,Yang,Doraiswamy,Rudzinski}. Although this problem
has been theoretically
\cite{Sircar,Heuchel,AYA,DUNNE,JCP8,CPL1,Myers,SMIT} and
experimentally \cite{Myers,SMIT,Gonzalez,Dunne,Dunne1} studied for
many years, some aspects are still unclear being necessary to
reach a better understanding about the behavior of the adsorbate
during the adsorption process of the mixture.

As in any adsorption process, a complete analysis of the behavior
of gas molecules under the influence of an adsorbent requires the
knowledge of the forces of molecular interactions
\cite{Ruthven,steele,alison,zda}. In other words, the description
of real multicomponent adsorption requires to take into account
the effect of the lateral interactions between each species of the
mixture. An exact treatment of this problem, including ad-ad
interactions, is unfortunately not yet available and, therefore,
the theoretical description of adsorption relies on simplified
models \cite{Rudzinski}. One way of overcoming this complication
is to use Monte Carlo (MC) simulation method
\cite{binder1,binder2,binder3,Unge,JMC1}. MC technique is a
valuable tool for studying surface molecular processes, which has
been extensively used to simulate many surface phenomena including
adsorption \cite{nicholson}, diffusion \cite{difu}, reactions,
phase transitions \cite{patry}, etc.

In this line of work, a previous article was devoted to the study
of the adsorption of interacting binary mixtures on triangular
lattices \cite{SS16}. In Ref. [\cite{SS16}], Rinaldi et al.
obtained adsorption isotherms and differential heats of adsorption
corresponding to both species of the mixture, for different values
of the lateral interactions between the adsorbed species. An
unusual feature was observed when $(i)$ the lateral interaction
between A and B particles was different from zero, and $(ii)$ the
initial concentration of B particles was in the range
$\left[0.3,0.5\right]$. In these conditions, the A particles
adsorbing on the lattice expel the B adsorbed particles; then, the
partial A [B] coverage increases [decreases]. During this regime
the number of desorbed particles is greater than the number of
adsorbed particles which results in a decreasing of the total
coverage that occurs for a wide range in values of chemical
potential where the slope of the adsorption isotherm is negative.
The behavior of the system was fully explained through the
analysis of the phase diagrams for order-disorder transitions
occurring in the adsorbed layer.

Because the structure of lattice space plays a fundamental role in
determining the statistics of mixtures, it is of interest and of
value to inquire how a specific lattice structure influences the
main thermodynamic properties of adsorbed mixtures. In this
context, the objectives of the present paper are (1) to extend the
previous work to square lattices using the same techniques
developed in Ref. [\cite{SS16}] and (2) to study the effect of the
lattice structure on the adsorption of interacting binary
mixtures. For this purpose, multicomponent gases adsorbed on
square lattices are studied by using grand canonical ensemble MC
simulation. The process was analyzed by following total and
partial adsorption isotherms as well as differential heats of
adsorption corresponding to both species of the mixture. The
detailed behavior of these quantities will be shown to be directly
related to the phase diagrams of the system.

The outline of the paper is as follows. In Sec. II we describe the
lattice-gas model and the simulation scheme. In Sec. III we
present the MC results. Finally, the general conclusions are given
in Sec. VI.

\section{Lattice-Gas Model and Monte Carlo Simulations}

The adsorptive surface is represented by a two-dimensional square
lattice of $M = L \times L$ adsorption sites, with periodic
boundary conditions. The substrate is exposed at a temperature $T$
to an ideal gas phase consisting of a binary mixture of particles
A and B with chemical potentials $\mu_A$ and $\mu_B$,
respectively. Particles can be adsorbed on the lattice with the
restriction of at most one adsorbed particle per site and we
consider a nearest-neighbor ($NN$) interaction energy $w_{XY}$ (X,
Y = A, B) among them. The adsorbed phase is then characterized by
the Hamiltonian:
\begin{eqnarray}
H &=& \frac{1}{2} \sum _{i} ^{M} \sum _{l\in \left\{ NN ,
i\right\} } \left[ w_{AA} \delta _{c_i , c_l , 1}  +
w_{BB} \delta _{c_i , c_l, -1} +    w_{AB} \left( \delta _{ c_i , 1} \delta _{c_l , -1 } + \delta _{c_i , -1} \delta_{c_l, 1} \right)   \right] + \nonumber \\
& & + \epsilon _0 \sum _i ^{M} \left( \delta _{c_i , 1}+\delta
_{c_i , -1}\right) - \sum _i ^M \left( \mu _A \delta _{ c_i,1} +
\mu _B \delta _{c_i,-1} \right)
\end{eqnarray}
where $c_i$ is the occupation number of site $i$ ($c_i = 0$ if
empty; $c_i = 1$ if occupied by A and $c_i = -1$ if occupied by
B); $l\in \left\{ NN , j\right\}$ runs on the four $NN$ sites of
site $i$; the $\delta$'s are Kronecker delta functions and
$\epsilon _0$ is the interaction energy between a monomer (type A
or B) and a lattice site. In this contribution, the chemical
potential of one of the components is fixed throughout the process
($\mu _B = 0$), while the other one ($\mu _A$) is variable, as it
is usually assumed in studies of adsorption of gas mixtures
\cite{Tovbin}. In the actual implementation of the model $\epsilon
_0$ was set equal to zero, without loss of any generality.

With respect to the computational simulations, the Monte Carlo
procedure used has been discussed in detail in Ref. [\cite{SS16}]
and need not be repeated here. In this case, the first $10^6$
Monte Carlo steps (MCS) were discarded to allow equilibrium, while
the next $10^6$ MCS were used to compute averages.

\section{Results and discussion}

The computational simulations have been developed for square $L
\times L$ lattices, with $L=96$, and periodic boundary conditions.
With this lattice size we verified that finite-size effects are
negligible. Note, however, that the linear dimension $L$ has to be
properly chosen such that the adlayer structure is not perturbed.

In order to understand the basic phenomenology, we consider in the
first place the case of single-gas adsorption. This was achieved
by making $\mu _B \to -\infty$. Fig. 1 shows the behavior of the
adsorption isotherms and the differential heats of adsorption for
different strengths of repulsive interparticle interactions. As
expected, we obtain the well-known Langmuir isotherm passing
through the point ($\mu _A/k_BT = 0$, $\theta _A = 1/2$) when
$w_{AA}/k_BT = 0$ (being $k_B$ the Boltzmann constant). Two
features, which are useful for the analysis of mixed-gas
adsorption, are worthy of comment: (a) as the $NN$ repulsive
interaction is increased, the coverage at zero chemical potential
decreases and asymptotically approaches $\theta _A = 0.226$; (b)
as the repulsive $NN$ interaction passes a critical value $w_c
/k_BT \approx 1.763$
 \cite{Kramers,yeomans}, a plateau develops in the isotherm at
$\theta _A = 1/2$ indicating the appearance of $c(2 \times 2)$
ordered phase on the surface. In what follows, we consider
mixed-gas adsorption but keeping species B at a fixed value of the
chemical potential $\mu _B/k_BT =0$. In addition, we have
considered $k_BT = 1$ for simplicity, without any lost of
generality.

We start with the case of a binary mixture in the presence of
repulsive lateral interactions between the particles. The effect
of AA interactions is depicted in Fig. 2, where $w_{AA}/k_BT \neq
0$, $w_{AB}/k_BT = 0$ and $w_{BB}/k_BT = 0$. We have plotted the
partial (a) and total (b) adsorption isotherms, and the
differential heats of adsorption corresponding to the species A
(c) and B (d). It can be observed in Fig. 2 (a) and (b), that the
initial coverage takes the same value $\theta=0.5$ for all
isotherms; this behavior can be explained as follows: for $\mu
_A/k_BT \to - \infty$ the A particle coverage is zero while the B
particles are randomly distributed on the lattice with $\theta_B$
given by the Langmuir isotherm
$\theta_B=\exp(\mu_B/k_BT)/[1+\exp(\mu_A/k_BT)+\exp(\mu_B/k_BT)]$,
which for $\mu_B/k_BT=0$ is $\theta=\theta_B=1/2$.

As discussed above, for a lattice gas of interacting monomers
adsorbed on a square surface, there exists a critical interaction
($w_c/k_BT \approx 1.763$) corresponding to a phase transition in
the adsorbate. The nearest-neighbor coupling, $w$, determines the
character of the phase transition: (i) if $w < 0$ (attractive
case) the system exhibits a first-order phase transition, and (ii)
for $w > 0$ (repulsive case) a continuous order-disorder phase
transition occurs in the adsorbate, which is observed as a clear
plateau in the adsorption isotherms. Following this behavior, a
well-defined and pronounced step appears in the partial isotherms
as the interaction $w_{AA}/k_BT$ is increased. Thus, the
interaction between A molecules determines a $c(2 \times 2)$
ordered phase for such particles. Therefore, the A isotherm
presents a plateau at half coverage [see Fig. 2(a)]. At
equilibrium, the B particles occupy half of the empty sites, and
the corresponding B isotherm presents a plateau at
$\theta_B=0.25$; this behavior is a consequence of the excluded
volume but is not due to the interactions. The total isotherm
[Fig. 2(b)] is the sum of the partial isotherms, then the plateau
appears at $\theta=0.75$.

In Fig. 2(c), the differential heat of adsorption $q_A$
corresponding to the A species is plotted versus $\theta_A$. The
behavior of the curves can be explained by analyzing two different
adsorption regimes: (i) for $0.5 < \theta < 0.75$ ($0 < \theta_A <
0.5$), the ad-molecules avoid $NN$ occupancy which produces $q_A
\approx 0$ (as for $w_{AA}/k_BT = 0$), and (ii) for $0.75 < \theta
< 1$ ($0.5 < \theta_A < 1$), the adsorption of one more molecule
involves an increment $cw_{AA}$ in the energy of the system, where
$c$ is the lattice connectivity (in this case, $c=4$). The maximum
in $q_A$ for $\theta \to 0.75^-$ corresponds to the critical
coverage at which a dramatic change of order takes place in the
system (the system passes from the disordered to the ordered phase
\cite{PHYSA3}). A similar situation occurs for the minimum in
$q_A$ at $\theta \to 0.75^+$.

The adsorption of the A species induces an interesting behavior in
the B isotherm, which also exhibits well-defined steps although
the B particles do not interact neither with B particles nor with
A particles ($w_{AB}/k_BT = 0$ and $w_{BB}/k_BT = 0$). This
behavior is a consequence of the excluded volume but is not due to
the interactions. Then, as it is expected, $q_B$ is strictly zero
over all the coverage range [see Fig. 2 d)].

We now continue the study of the effect of AA interactions with
$w_{AB}/k_BT = 0$ and $w_{BB}/k_BT = 2$ (Fig. 3), therefore
introducing BB interactions; $w_{AB}/k_BT = 2$ and $w_{BB}/k_BT =
0$ (Fig. 4) introducing interspecies interactions (and removing BB
interactions) and, finally, we analyze $w_{AB}/k_BT = 2$ and
$w_{BB}/k_BT = 2$ (Fig. 5) where all interactions are present.

In the case of Fig. 3, the behavior of the  partial [Fig. 3 a)]
and total [Fig. 3 b)] isotherms and the differential heat of
adsorption $q_A$ [Fig. 3 c)] is similar to those in Fig. 2. The
main difference is associated to the value of $\theta _B$ at low
pressures (which is lower than 0.5), and to the behavior of $q_B$
that is not zero any more.

Clearly, for $w_{AA}/k_BT = 0$, A particles are distributed at
random and B particles start from a low coverage (close to
$0.226$, as explained in Fig. 1) which rapidly decreases as more A
particles are adsorbed. Therefore as the total coverage increases,
B particles interact with each other less frequently and $q_B$
increases steadily. However, as $w_{AA}/k_BT$ becomes sufficiently
high so that an ordered phase is formed, a sudden increase in
$q_B$ is produced due to a sudden increase in the screening effect
between B particles produced by A particles.

An unusual feature is observed in the case of Fig. 4: as $\mu
_A/k_BT$ increases and A particles start to adsorb, B particles
are displaced from the surface so that the total A + B coverage
decreases and shows a local minimum [Fig. 4 b)]. This effect,
which has been previously called {\em mixture effect} \cite{SS16},
can be explained as follows: as the coverage of A particles is
sufficiently high so that AB interactions occur, the repulsive
character of $w_{AB}$ leads to more B particles being displaced
from the surface than A particles being adsorbed on the surface.

The mixture effect is clearly reflected in the behavior of $q_A$
[Fig. 4 c)] and $q_B$ [Fig. 4 d)]. In fact, for $0 < \theta _A <
1/2$, A particles do not interact with other A or B molecules and,
consequently, $q_A = q_B = 0$ (see insets). The $c(2 \times 2)$
phase of A particles starts to develop and is completed at $\theta
_A = 1/2$. For $1/2 < \theta _A$, A particles fill the vacancies.
In this regime, the coverage of B particles tends to zero and does
not perturb significantly the adsorption of the A species. The
important fluctuations in $q_B$ are clear signals that the number
of B particles is practically zero.

Our simulations show how the competition between two species in
presence of repulsive mutual interactions reinforces the
displacement of one species by the other and leads to the presence
of the mixture effect. To complete this analysis, it is
interesting to note that the mixture effect [also called
adsorption preference reversal (APR) phenomenon] has also been
observed for methane-ethane mixtures \cite{AYA,DUNNE,JCP8,CPL1}.
The rigorous results presented in Ref. [\cite{CPL1}] showed that,
in the case of methane-ethane mixtures, the APR is not a
consequence of the existence of repulsive interactions between the
ad--species, but it is the result of the difference of size (or
number of occupied sites) between methane and ethane. A similar
scenario has been observed for different mixtures of linear
hydrocarbons in silicalite \cite{SMIT}, carbon nanotube bundles
\cite{Jiang} and metal-organic frameworks \cite{Jiang1}.

In the case of Fig. 5, the presence of repulsive BB interactions
results in a initial coverage of B particles close to $0.226$.
This small fraction of B molecules does not perturb significantly
the adsorption of the A species and the mixture effect disappears.
This finding indicates that, in addition to the requirements of
repulsive lateral interactions between the two species, the amount
of particles on the surface is essential for the existence of the
mixture effect. The rest of the figure can be understood following
the arguments given above.

In the following we study the effects of variable AB interactions
as shown in Figs. 6 to 9. We start with the case where
$w_{AA}/k_BT = w_{BB}/k_BT = 0$ (Fig. 6). Here no ordered phases
are formed and for sufficiently high $w_{AB}/k_BT$ an important
mixture effect appears. Let us choose for our analysis the curves
corresponding to $w_{AB}/k_BT = 1$. B coverage is initially 0.5.
As A molecules are adsorbed B molecules are eliminated from the
surface in such a way that $\theta$ decreases. At the same time
$q_A$ starts at a low value and increases rapidly (with a
bivaluated behavior) tending to 0 as $\theta \to 1$, while $q_B$
starts near 0 and tends to $4w_{AB}$ as $\theta \to 1$. As soon as
BB interactions are added, the starting B coverage is $\sim 0.226$
and the mixture effect disappears (Fig. 7). In the presence of AA
and AB interactions (Fig. 8), $w_{AB}/k_BT = 0.5$ is sufficiently
high for the appearance of ordered phases for A molecules, so that
the resulting behavior is similar to that of Fig. 4. The inclusion
of the three interactions (Fig. 9), only produces the elimination
of the mixture effect, compared to Fig. 8, due to the low initial
B coverage.

The effect of variable BB interactions is discussed in Figs. 10 to
13. Different cases were considered: $w_{AA}/k_BT = w_{AB}/k_BT =
0$ (Fig. 10); $w_{AA}/k_BT = 0$ and $w_{AB}/k_BT = 2$ (Fig. 11);
$w_{AA}/k_BT = 5$ and $w_{AB}/k_BT = 0$ (Fig. 12) and $w_{AA}/k_BT
= 5$ and $w_{AB}/k_BT = 2$ (Fig. 13). Given the value of the
parameters in Fig. 10, neither the coverage of A [Fig. 10 a)] nor
the differential heat of adsorption [Fig. 10 c)] are affected by
BB interactions. As discussed in Fig. 1, the coverage of B at low
pressure starts at $0.5$ and decreases as $w_{BB}/k_BT$ increases
towards the limiting value of $0.226$ [see Fig. 10 a)]. However,
the coverage of B does not present any special features. Such a
special feature does indeed appear in the behavior of $q_B$ [Fig.
10 d)], which decreases steadily as $w_{BB}/k_BT$ increases below
a certain critical value, $w^c_{BB}/k_BT$ (being $w^c_{BB}/k_BT
\approx 2$), and increases above it. The explanation for this
behavior is that B particles adsorb more or less at random below
$w^c _{BB}/k_BT$, thereby allowing some BB interactions which
contribute to the decrease in the differential heat of adsorption.
Above $w^c_{BB}/k_BT$, B particles adsorb forming an ordered
structure so that BB interactions stop contributing and $q_B$
increases. The condition $w_{AB}/k_BT = 0$ restricts the
possibility of mixture effect.

In Fig. 11, the presence of AB interactions favors the
displacement of B particles and, consequently, the slopes of the B
partial isotherms are increased. On the other hand, as the initial
fraction of B particles is high ($0.3\le \theta^i_B \le 0.5$),
which occurs for small values of $w_{BB}/k_BT$ ($ 0 \le
w_{BB}/k_BT \le 0.5$), AB interactions lead to mixture effect. In
a wide range of coverage, $ 0.4 \le \theta \le 1$, the curves in
Fig. 11 are very similar between them (this is clearly visualized
in the case $q_A$ and $q_B$), which is indicative of the rapid
decreasing of the number of B particles on the surface.

In Fig. 12, the adsorption of A particles [Fig. 12 a)] follows a
unique isotherm and is independent of the strength of BB
interactions. Adsorption of B particles decreases at low A
coverage with increasing values of $w_{BB}/k_BT$ and tends to the
limiting value $\theta _B = 0.226$. However, at high A coverage
all isotherms tend to that corresponding to $w_{BB}/k_BT = 0$
because there are no $NN$ vacant sites in that range available for
the adsorption of B particles. This determines the total coverage
behavior shown in Fig. 12 b). The curves of $q_A$, shown in Fig.
12 c), present a very similar behavior between them. This is, one
marked jump appears, corresponding to the plateau in A isotherms.
In contrast, $q_B$ presents two types of behaviors [Fig. 12 d)].
Thus, at very low $w_{BB}/k_BT$ values (negligible interactions)
and very high $w_{BB}/k_BT$ values (above the critical value for
the formation of the ordered phase), it remains practically
constant, while at intermediate values it increases in line with
the total coverage. In the last case (Fig. 13), the adsorption
process can be easily understood: the B particles disappear for
low values of $\mu _A/k_BT$. Then, for higher $\mu _A/k_BT$'s, all
curves collapse on a unique curve.

\subsection{Phase diagrams}

In order to rationalize the results presented in previous section,
we will determine the temperature-coverage phase diagram
characterizing our system in the range of the parameters studied.
The curves will be obtained as a generalization of the well-known
phase diagram for a lattice-gas of repulsive monomers adsorbed on
a homogeneous square lattice, which is shown in Fig. 14.

Some of the exact properties of this system have been found,
especially by Onsager \cite{Onsager}. These are confined mostly to
the special condition $\theta =1/2$, but by symmetry, it can be
deduced that $\theta_c=1/2$, if a critical point exists, so this
is the most interesting value of $\theta$. Thus, the maximum of
the coexistence curve (occurring to $\theta = 1/2$) corresponds to
a critical value $k_B T_c /w_{AA} = \left[|2 \ln(\sqrt{2}-1)|
\right]^{-1} \approx 0.567$ \cite{Kramers,yeomans}. On the other
hand, zones I, II and III correspond to a disordered lattice-gas
state, a ordered state [$c(2 \times 2)$ phase], and a disordered
lattice-liquid state, respectively.

We start with the analysis of Fig. 13, where our model is almost
identical to a lattice-gas of one species. As indicated in Fig.
13, the existence of repulsive A–B interactions favors the
displacement of B particles. Once $\theta _B \approx 0$, which
occurs at $\mu _A / k_BT \approx 5$, the binary mixture is
equivalent to the square lattice-gas of one species. To
corroborate this affirmation, Fig. 15 shows the total isotherms of
Fig. 13 b) (symbols), in comparison with the adsorption isotherm
corresponding to a square lattice-gas of one species and $w_{AA}/
k_BT = 5$ (solid line). The analysis can be separated in two
parts: for $\theta < 1/2$, there exist B particles on the lattice
and, consequently, the curves of the mixture deviate from that
corresponding to one species. For $\theta > 1/2$, the coverage of
B particles is negligible and all curves collapse on a unique
curve. Then, the phase diagram characterizing a binary mixture
with the set of parameters of Fig. 13 is identical to that shown
in Fig. 14. The unique difference with the one-species phase
diagram is that the zone I now corresponds to an A-B mixture with
different proportions according to the value of $w_{BB}/ k_BT$.

As a basis for the analysis of the behavior of the system for
variable $w_{AB}/ k_BT$, we begin by considering one of the cases
of Fig. 13 (that corresponding to $w_{BB}/ k_BT = 0$), which is
characterized by a phase diagram as shown in Fig. 14. This
behavior is representative of systems with high values of $w_{AB}/
k_BT$ (in the range $w_{AB}/ k_BT \geq 2.0$), as is indicated in
Fig. 16 (see the plane corresponding to $w_{AB}/ k_BT = 2.0$). As
$w_{AB}/ k_BT$ is decreased, the maxima of the coexistence curves
in zone II shift to higher values of coverage\footnote{The figure
shows the upper part of each coexistence curve. A complete
analysis of the curve should require a more complex study
(percolation of phase, zero--temperature calculations, etc.),
which are out of the scope of the present paper.}. This can be
better visualized in Fig. 17 (solid circles, bottom axis), where
we plot the densities corresponding to the maxima of the
coexistence curves in zone II, $\theta _c$'s, as a function of
$w_{AB}/ k_BT$. The figure allows to analyze the behavior of a
binary mixture with $w_{BB}/ k_BT = 0$, variable $w_{AB}/ k_BT$
and $w_{AA}/ k_BT$ in the critical regime. The curves can be
understood according to the following reasoning. As it was
explained for high values of the AB lateral interaction, the zone
II corresponds to a phase of A particles. As $w_{AB}/ k_BT$
decreases, the phase is complemented with B particles, which are
randomly distributed in the empty sites of the structure. As a
typical example, we will analyze the case of $w_{BB}/ k_BT=
w_{AB}/ k_BT= 0$. In this case, the B particles, which are at
chemical potential $\mu _B / k_BT= 0$, occupy at random $1/2$ of
the empty sites of the phase\footnote{Note that $1/2$ of the empty
sites of the phase $c(2 \times 2)$ represents $1/4$ of the total
sites.}. Then, the phase corresponding to zone II is formed by a
structure of A particles (which occupy $1/2$ of the total sites)
complemented with B particles (which occupy at random $1/4$ of the
total sites). The resulting value of $\theta _c$ is $3/4$. The
rest of the points in Fig. 17 can be explained by similar
arguments.

We now turn to the effect of BB interactions (see Fig. 17, open
circles, upper axis). For this purpose, we set $w_{AB}/ k_BT = 0$
and $w_{AA}/ k_BT$ in the critical regime. The adsorption
properties corresponding to this case were discussed in Fig. 12.
We start the analysis with the case $w_{BB}/ k_BT = w_{AB}/ k_BT =
0$, where $\theta _C = 3/4$. As $w_{BB}/ k_BT$ is increased,
$\theta _C$ remains constant. In this case, the phase corresponds
to a $c(2 \times 2)$ phase of A particles complemented with a
partial coverage of B particles equal to $1/4$.

Figs. 14-17 allow to characterize the critical behavior of a
binary mixture adsorbed in a square lattice in the region of the
parameters studied in the present contribution. Namely:

\begin{itemize}

\item{} In Fig. 2, the total isotherms have a pronounced plateau
at $\theta = 3/4$ for strongly repulsive AA interactions, which
smoothes out already for $w_{AA}/ k_BT < 1$. This result is
reflected in Fig. 17.

\item{} Figs. 3 and 4 correspond to particular cases in Fig. 17.

\item{} Figs. 5 and 9 can be explained by combining the results in
Fig. 17.

\item{} Figs. 8, 12 and 13 were discussed in details above.

\item{} Finally, no phase transition develops in the system when
$k_BT/w_{AA} > 0.567$. This is clearly seen in Figs. 6, 7, 10 and
11, where a smooth behavior in the adsorption properties is
obtained.

\end{itemize}

\section{Conclusions}

Using Monte Carlo simulations, we have studied the adsorption of a
gas mixture of interacting particles A and B on homogeneous square
surfaces. A variety of behaviors arise due to the formation of
different ordered structures in the adlayer for different values
of the lateral interactions among adsorbed particles. The analysis
of partial and total adsorption isotherms and differential heats
of adsorption provides a detailed understanding of the adsorption
process. This study yields to the construction of a phase-diagram
which allows to understand the critical behavior of the system.

An interesting feature of this work is the occurrence of a minimum
in the global adsorption isotherm. This singularity appears as the
initial fraction of B particles on the surface is high ($0.4 \leq
\theta^i _B \leq 0.5$) and $w_{AB}/k_BT > 0$. This effect might be
interesting due to the fact that seems counterintuitive to see a
negative slope (therefore a minimum) in the total coverage.
Nevertheless, the phenomenon is because the A particles adsorbing
in the lattice expel the B particles at a higher ratio; then the
partial A [B] coverage increases [decreases] (partial isotherms
must cross). In the above regime, the desorbed B particles are
more than the adsorbed A particles, therefore we see a total
coverage ($\theta$) with a negative slope followed by a minimum.

The computational technique used here has proven to be very
powerful tool for these kind of lattice-gas models and many other
systems in a much wider scope of sciences, allowing the
interpretation of many experimental results without heavy or
time-consuming calculations.

\section{ACKNOWLEDGMENTS}

This work was supported in part by CONICET (Argentina) under
project number PIP 112-200801-01332; Universidad Nacional de San
Luis (Argentina) under project 322000; Universidad Tecnol\'ogica
Nacional, Facultad Regional San Rafael (Argentina) under projects
PQPRSR 858 and PQCOSR 526 and the National Agency of Scientific
and Technological Promotion (Argentina) under project 33328 PICT
2005.

\newpage

\section*{Figure Captions}

\noindent Fig. 1: Adsorption isotherms for the single-gas
adsorption of A particles onto the surface showing the effect of
lateral AA interactions.

\noindent Fig. 2: Mixed-gas adsorption on a square lattice: (a)
adsorption isotherms for A and B particles; (b) total adsorption
isotherms; (c) differential heat of adsorption for A particles and
(d) differential heat of adsorption for B particles. Effect of AA
interactions: $w_{AA}/k_BT \geq 0$, $w_{BB}/k_BT = 0$ and
$w_{AB}/k_BT = 0$.

\noindent Fig. 3: As Fig. 2 for $w_{AA}/k_BT \geq 0$, $w_{BB}/k_BT
= 2$ and $w_{AB}/k_BT = 0$.

\noindent Fig. 4: As Fig. 2 for $w_{AA}/k_BT \geq 0$, $w_{BB}/k_BT
= 0$ and $w_{AB}/k_BT = 2 $.

\noindent Fig. 5: As Fig. 2 for $w_{AA}/k_BT \geq 0$,
 $w_{BB}/k_BT = 2 $ and $w_{AB}/k_BT = 2 $.

\noindent Fig. 6: Mixed-gas adsorption on a square lattice: (a)
partial adsorption isotherms for A and B particles; (b) total
adsorption isotherms; (c) differential heat of adsorption for A
particles and (d) differential heat of adsorption for B particles.
Effect of AB interactions: $w_{AA}/k_BT = 0$,
 $w_{BB}/k_BT = 0$ and $w_{AB}/k_BT \geq 0$.

\noindent Fig. 7: As Fig. 6 for $w_{AA}/k_BT = 0$,
 $w_{BB}/k_BT = 2$ and $w_{AB}/k_BT \geq 0$.

\noindent Fig. 8: As Fig. 6 for $w_{AA}/k_BT = 5 $,
 $w_{BB}/k_BT = 0$ and $w_{AB}/k_BT \geq 0$.

\noindent Fig. 9: As Fig. 6 for $w_{AA}/k_BT = 5 $,
 $w_{BB}/k_BT = 2$ and $w_{AB} /k_BT \geq 0$.

\noindent Fig. 10: Mixed-gas adsorption on a square lattice: (a)
partial adsorption isotherms for A and B particles; (b) total
adsorption isotherms; (c) differential heat of adsorption for A
particles and (d) differential heat of adsorption for B particles.
Effect of BB interactions: $w_{AA}/k_BT = 0$,
 $w_{BB}/k_BT \geq 0$ and $w_{AB}/k_BT = 0$.

\noindent Fig. 11: As Fig. 10 for $w_{AA}/k_BT = 0$,
 $w_{BB}/k_BT \geq 0$ and $w_{AB}/k_BT = 2$.

\noindent Fig. 12: As Fig. 10 for $w_{AA}/k_BT = 5$,
 $w_{BB}/k_BT \geq 0$ and $w_{AB}/k_BT = 0$.

\noindent Fig. 13: As Fig. 10 for $w_{AA}/k_BT = 5$,
 $w_{BB}/k_BT \geq 0$ and $w_{AB}/k_BT = 2$.

\noindent Fig. 14: Temperature-coverage phase diagram
corresponding to a lattice-gas of repulsive monomers ($w_{AA}/k_BT
> 0$) adsorbed on a homogeneous square lattice.

\noindent Fig. 15: Comparison between the total isotherms in Fig.
13 b) (symbols) and the one corresponding to a square lattice-gas
of one species with $w_{AA}/k_BT = 5$ (solid line). The inset
shows a snapshot of the $c(2 \times 2)$ phase.

\noindent Fig. 16: Effect of the lateral interactions between A-B
particles, $w_{AB}/k_BT$, on the temperature-coverage phase
diagram corresponding to a binary mixture with $w_{BB}/k_BT = 0$
and $w_{AA}/k_BT$ in the critical regime.

\noindent Fig. 17: Densities corresponding to the maxima of the
coexistence curves in zone II as a function of $(i)$ $w_{AB}/k_BT$
(solid circles, bottom axis) and $(ii)$ $w_{BB}$ (open circles,
upper axis). In case $(i)$ [$(ii)$], $w_{BB}/k_BT = 0$
[$w_{AB}/k_BT = 0$] and $w_{AA}/k_BT$ is chosen in the critical
regime.

\newpage



\end{document}